\documentclass[12pt]{article}
\usepackage{amssymb}
\usepackage{graphicx}
\usepackage[cp866]{inputenc}
\begin{document}
 \begin{center}

{\bf Positronia' clouds in Universe}

\vspace{2mm}

I.M. Dremin\footnote{e-mail: dremin@lpi.ru}\\

{\it Lebedev Physical Institute, Moscow, Russia}

\end{center}


\vspace{1mm}

\begin{abstract}
The intense emission of 511 keV photons from the Galactic center and within 
terrestrial thunderstorms is attributed to the formation of parapositronia' 
clouds. Unbound electron-positron pairs and positronia can be created by strong 
electromagnetic 
fields produced in interactions of electrically charged objects, in particular,    
in collisions of heavy nuclei. Kinematics of this process favours abundant 
creation of the unbound electron-positron pairs with very small masses and the 
confined parapositronia states which decay directly to two 511 keV quanta. 
Therefore we propose to consider interactions of electromagnetic fields
of colliding heavy ions as a source of low-mass pairs which can transform
to 511 keV quanta. Intensity of their creation is enlarged 
by the factor Z$^4$ (Z is the electric charge of a heavy ion) compared 
to protons with Z=1. These processes are especially important at very high
energies of nuclear collisions because their cross sections increase
proportionally to cube of the logarithm of energy and can even exceed the
cross sections of strong interactions which may not increase faster than
the squared logarithm of energy. Moreover, production of extremely low-mass 
$e^+e^-$-pairs in ultraperipheral nuclear collisions is strongly enhanced due 
to the Sommerfeld-Gamow-Sakharov (SGS) factor which accounts for mutual Coulomb
attraction of non-relativistic electrons to positrons in case of low pair-masses. 
This attraction may lead to their annihilation and, therefore, to the increased 
intensity of 511 keV photons. It is proposed to confront the obtained results
to forthcoming experimental data at NICA collider.                 
\end{abstract}

\vspace{1mm}


\vspace{1mm}

\section{Introduction}

The intense emission of 511 keV gamma-rays from the Galactic center 
\cite{sieg} and within active terrestrial thunderstorms \cite{abc} has been 
observed. It seems quite natural to attribute both observations to the formation 
of clouds of parapositronia, each decaying into two gamma-quanta with energy
511 keV. Moreover, one is tempted to relate these facts to the well known in 
experimental particle physics excess \cite{skl} of soft dileptons 
and photons over their calculated supply from all well established sources.

Characteristic gamma-ray emission from the Galaxy's interstellar medium
with a line at 511 keV has been observed with the spectrometer SPI on ESA's 
INTEGRAL observatory \cite{sieg}. It registered bright emission from an 
extended bulge-like region, while emission from the disk is faint.

Terrestrial gamma-ray flashes are usually observed by spacecraft in low-Earth 
orbit, but have also been observed by aircrafts \cite{abc} and on the ground
\cite{chub1, chub2, chil} during active thunderstorms. Their energy spectra 
are consistent with a source mostly composed of positron annihilation 
gamma-rays, with a prominent 511 keV line clearly visible in the data.

It is reasonable to assume that electromagnetic fields are responsible for
this effect. The outcome of a collision of two charged objects (two nuclei,
in particular) is especially soft if they do not come 
very close to one another but pass at large distances (impact parameters) 
and only their electromagnetic fields interact. These processes are named 
ultraperipheral because of the large spatial extension of electromagnetic 
forces compared to the much lower range of strong nuclear interactions. 
It is shown here that very soft dielectrons and photons are produced in these 
grazing collisions \cite{dr1, dr2}.

\section{Sources of positrons and positronia}

The main problem of the hypothesis about positronia' clouds charged for
emission of the 511 keV photon peak is the rate of parapositronium
production. Electrons are very abundant as main ingredients of atoms
while positrons must be produced in some interactions. Electromagnetic and weak
forces must be responsible for their production. Parapositronia and unbound
electron-positron pairs can be created by electromagnetic fields of any moving 
charged objects with energy exceeding the threshold of their production.
Weak forces are active in radioactive nuclear decays and decays of unstable
particles created in strong interactions.

Astrophysicists analyze intensively the objects in the Universe which could be
responsible for these effects. The long list of candidates has been proposed.
It includes radioactive decay of unstable isotopes produced in nucleosynthesis 
sources throughout the Galaxy \cite{dieh}, accreting binary systems 
(microquasars) \cite{takh}, old neutron stars (former pulsars) \cite{isto}, 
various sources inside the supermassive black hole in our Galaxy's center 
\cite{dogi, cai, takh}, dark matter decay or annihilation \cite{farz, cai}, as 
dark matter would be gravitationally concentrated in the inner Galaxy etc.
Quantitative estimates of their contributions leave considerable uncertainties.
The puzzle still remains, and no conclusive candidate has emerged.

Other complexities arise because, once ejected from their sources, the positrons 
may travel far from their origins to the surrounding interstellar gas. 
Kinetic equations are often used to show that processes of Coulomb collisions 
are effective enough to cool down these relativistic positrons and to thermalize 
them before their annihilation. Positrons slow down to energies of a few eV 
and annihilation may occur. 

Here, we do not enter in the detailed discussion of these problems but
propose the mechanism of direct electromagnetic creation of parapositronia and 
low-mass electron-positron pairs in ultraperipheral collisions of heavy ions.
The results can be confronted to forthcoming experimental data of NICA
collider and their relevance to the above findings established.
 
\section{Ultraperipheral nuclear collisions at NICA collider}

Production of $e^+e^-$-pairs in electromagnetic fields of colliding heavy ions
was first considered by Landau and Lifshitz in 1934 \cite{lali}. It was
shown that the total cross section of this process rapidly increases with
increasing energy $E$ as $\ln ^3E$ in asymptotics. This is still the
strongest energy dependence in particle physics. The cross sections of
strong nuclear interactions can not increase faster than $\ln ^2E$
according to the Froissart theorem \cite{froi}. Moreover, the numerical factor 
$Z^4\alpha ^4$ compensates in the total cross section the effect of the small 
electromagnetic coupling $\alpha $ for heavy ions with large charge $Ze$. 
Therefore, the ultraperipheral production of $e^+e^-$-pairs (as well as 
$\mu ^+\mu ^-$ etc.) in ion collisions can become the dominant mechanism at 
very high energies \cite{dr3}. It is already widely studied at colliders. At 
the same time, it is desirable to learn these processes at conditions closest to
astrophysics requirements. In particular, the energies of NICA collider
are above the threshold of dielectrons production but lower than
the threshold for dimuons \cite{ijmp}. That is why we demonstrate here the
properties of positronia (and 511 keV photons) and dielectrons just
in this energy interval from 5 GeV to 11 GeV per nucleon in the center
of mass system. We show that 
the heuristic knowledge of these processes is helpful in understanding some 
astrophysical phenomena as well.

Abundant creation of pairs with rather low masses is the typical feature of
ultraperipheral interactions \cite{dr1}. Unbound lepton pairs and 
parapositronia are produced in grazing collisions of interacting ions where 
two photons from their electromagnetic clouds interact. Two-photon fusion 
production of lepton pairs has been calculated with both the equivalent photon 
approximation proposed in \cite{wei, wil} and via full lowest-order QED 
calculations \cite{rac, bgms} reviewed recently in \cite{drufn}.
According to the equivalent photon approximation, the spectra of dileptons
created in ultraperipheral collisions
can be obtained from the general expression for the total cross section
\begin{equation}
\sigma _{up}(X)=
\int dx_1dx_2\frac {dn}{dx_1}\frac {dn}{dx_2}\sigma _{\gamma \gamma }(X).
\label{e2}
\end{equation}
Feynman diagrams of
ultraperipheral processes contain the subgraphs of two-photon interactions
leading to production of some final states $X$ (e.g., $e^+e^-$ pairs).
These blobs can be represented by the cross sections of these processes.
Therefore, $\sigma _{\gamma \gamma }(X)$ in (\ref{e2}) denotes the total cross
section of production of the state $X$ by two photons from the electromagnetic
clouds surrounding colliding ions and $dn/dx_i$ describe the densities of
photons carrying the share $x_i$ of the ion energy.

The~distribution of equivalent photons with a fraction of the nucleon energy
$x$ generated by a moving nucleus with the charge $Ze$ can be denoted as
\begin{equation}
\frac {dn}{dx}=\frac {2Z^2\alpha }{\pi x}\ln \frac {u(Z)}{x}
\label{flux}
\end{equation}
if integrated over the transverse momentum up to some value (see, e.g.,
\cite{blp}). The physical meaning of the ultraperipherality
parameter $u(Z)$ is the ratio of the maximum adoptable transverse momentum
to the nucleon mass as the only massless parameter of the problem.
Its value is determined by the form factors of colliding ions
(see, e.g., \cite{vyzh}).
It is clearly seen from Eq. (\ref{flux}) that soft photons with small
fractions $x$ of the nucleon energy dominate in these fluxes.

The cross section $\sigma _{\gamma \gamma }(X)$ usually inserted
in (\ref{e2}) in case of creation of the unbound dielectrons $X=e^+e^-$
is calculated in the lowest order perturbative approach and
looks \cite{blp,brwh} as
\begin{equation}
\sigma _{\gamma \gamma }(X)=\frac {2\pi \alpha ^2}{M^2}
[(3-v^4)\ln \frac {1+v}{1-v}-2v(2-v^2)],
\label{mM}
\end{equation}
where $v=\sqrt {1-\frac {4m^2}{M^2}}$ is the velocity of the pair components 
in the pair rest system, $m$ and $M$ are the electron and dielectron masses,
correspondingly. The cross section tends to 0 at the threshold of pair
production $M=2m$ and decreases as $\frac {1}{M^2}\ln M$ at very large $M$.

The distribution of masses $M$ of dielectrons is obtained after inserting
Eqs (\ref{flux}), (\ref{mM}) into (\ref{e2}) and leaving free one integration
there. One gets \cite{dr1}
\begin{equation}
\frac {d\sigma }{dM}=\frac {128 (Z\alpha )^4}{3\pi M^3}
[(1+\frac {4m^2}{M^2}-\frac {8m^4}{M^4})
\ln \frac {1+\sqrt {1-\frac {4m^2}{M^2}}}{1-\sqrt {1-\frac {4m^2}{M^2}}}-
(1+\frac {4m^2}{M^2})\sqrt {1-\frac {4m^2}{M^2}}]
\ln ^3\frac {u\sqrt {s_{nn}}}{M},
\label{sM}
\end{equation}
where $\sqrt {s_{nn}}$ is the c.m.s. energy per a nucleon pair. The dielectron
distribution (\ref{sM}) is shown in Fig. 1 for three NICA energies ranging 
from 11 GeV to 8 GeV and 6.45 GeV per nucleon. The parameter $u$=0.02 has been
chosen in accordance with its value obtained in Ref. \cite{vyzh} where
careful treatment of nuclei form factors is done. The total cross section of
ultraperipheral production of unbound pairs is
\begin{equation}
\sigma (ZZ(\gamma \gamma )\rightarrow ZZe^+e^-)=\frac {28}{27}
\frac {Z^4\alpha ^4}{\pi m^2}\ln^3\frac {u^2s_{nn}}{4m^2}.
\label{vz}
\end{equation}

\begin{figure}

\centerline{\includegraphics[width=16cm, height=14cm]{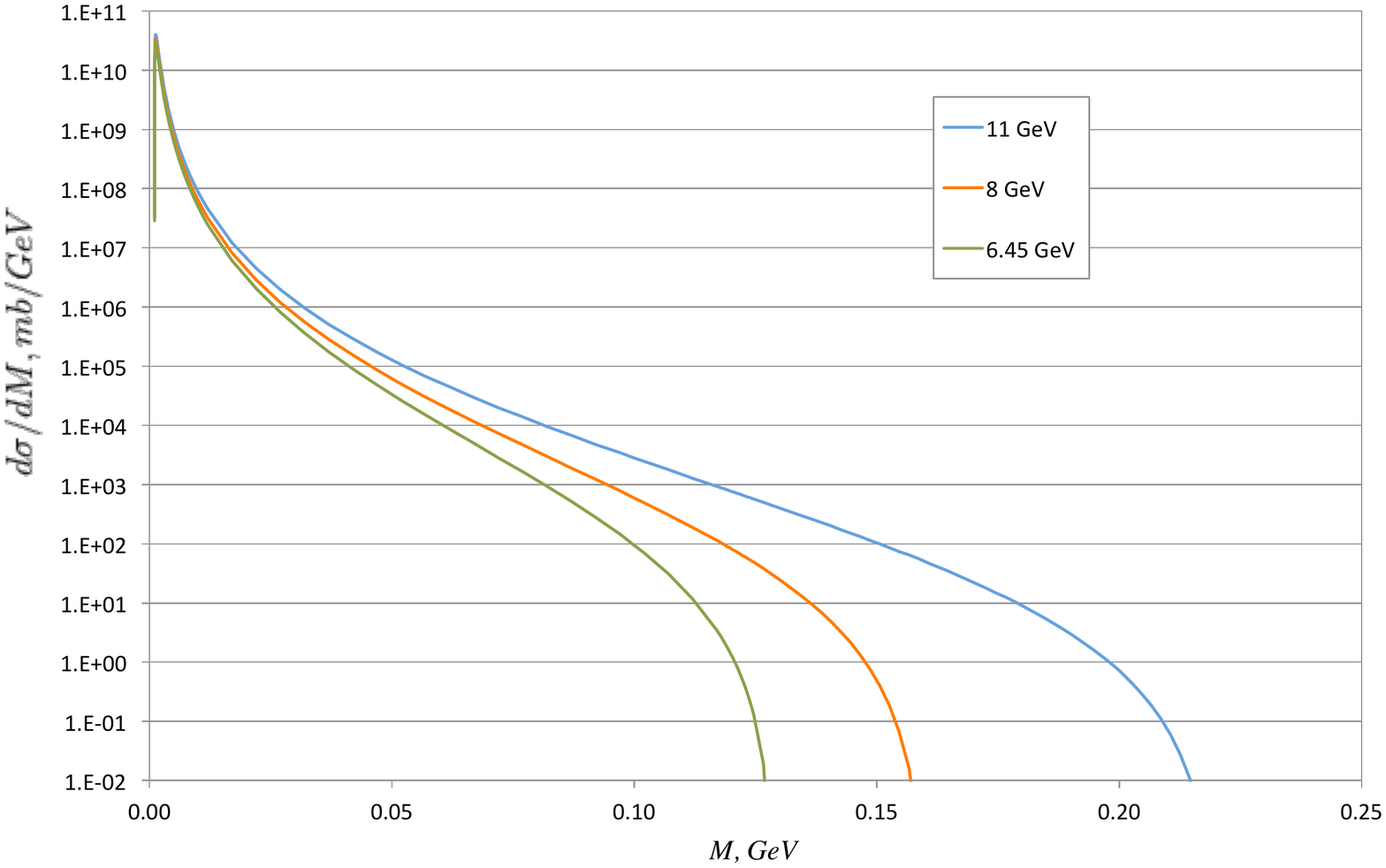}}

Fig. 1. The distribution of masses of dielectrons produced in
ultraperipheral collisions at NICA energies $\sqrt {s_{nn}}$=11 GeV 
(blue, upper), 8 GeV (red, middle), 6.45 GeV (green, lower).
\end{figure}

The sharp peak at low masses $M$ in Fig. 1 demonstrates the most important
feature of ultraperipheral processes - the abundant production of soft 
dielectrons with masses of the order of several electron masses $m=0.511$ MeV 
(effectively, less than 5 MeV). Moreover, the non-perturbative effects 
drastically enlarge production of unbound pairs with extremely low masses.

The perturbative expression for the cross section $\sigma _{\gamma \gamma }(X)$
of (\ref{mM})  can be generalized to include the non-perturbative effects
crucial near the pair production threshold $M=2m$. It happens to be possible for Coulomb
interaction governing the behavior of the components of a pair.
At the production point, the components of pairs with low masses close to 2$m$
move very slowly relative to one another. They are strongly influenced by
the attractive Coulomb forces. In the non-relativistic limit, these
states are transformed by mutual interactions of the components to effectively
form a composite state whose wave function is a solution of the relevant
Schroedinger equation. The normalization of Coulomb wave functions plays an
especially important role at low velocities. It differs from the normalization
of free motion wave functions used in the perturbative derivation of
Eq. (\ref{mM}). 

 The amplitude $R_C$ of the process $\gamma\gamma\to e^+e^-$ with account of 
the interaction between leptons is connected to the amplitude $R_0$ without the
final state interaction by the relation 
\begin{equation}
R_C=\int \Psi_f(r)R_0(r)d^3r
\label{rc}
\end{equation}
where $\Psi_f(r)$ is the wave function for bound (parapositronium) 
or unbound lepton pairs in the coordinate representation. 

For lepton pairs in $S$-state (the orbital momentum $l$=0) the
characteristic distances of the pair production are $~1/m$,  whereas the 
Coulomb interaction between leptons acts over the much larger distances 
$~1/{m\alpha}$ in the bound state production and $~1/k$ for the unbound 
states where $k$ is the relative momentum. Therefore, the wave function can be
considered as constant in (\ref{rc}) and one gets
\begin{equation}
 R_C= \Psi_{kS}(r=0)\int R_0(r)d^3r= \Psi_{kS}(r=0) R_0(p=0).
 \label{rc0}
\end{equation}
This relation is valid not only for bound states, but also 
for the creation of the unbound lepton pairs if $kr_s\ll 1$. 
Such factorization of matrix elements has been widely used in the dimesoatoms 
production. It is useful  for any process where the  
characteristic distances of pair production and of final state interactions 
are substantially different.
The normalization of the unbound pair wave function reads \cite{llqm}
\begin{equation}
|\psi_{kS}(\vec r=0)|^2 =\frac{\pi\xi}{sh(\pi\xi)}e^{\pi\xi}
=\frac{2\pi\xi}{1-e^{-2\pi\xi}};~~~\xi=\frac{2\pi\alpha m}{k}.
\label{psi}
\end{equation}
This is the widely used Sommerfeld-Gamow-Sakharov (SGS) factor 
\cite{som, gam, somm, sakh} which unites the non-perturbative and perturbative 
matrix elements. It results in the so-called "$\frac {1}{v}$-law" of the 
enlarged outcome of the reactions with extremely low-mass pairs produced. 
This factor is described in the standard
textbooks on non-relativistic quantum mechanics (see, e.g., \cite{llqm})
and used in various publications (e.g., \cite{baier, ieng, cass, arko}). 
The Sakharov recipe of its account for production of $e^+e^-$-pairs desctibed 
in \cite{sakh} consists in direct multiplication of the differential 
distribution of Eq. (\ref{sM}) by the SGS-factor written as
\begin{equation}
T=\frac {2\pi \alpha}{v(1-\exp (-2\pi \alpha/v))}.
\label{sgs}
\end{equation}
It enhances the contribution of the low-mass (low-$v$) pairs.
Thus the proper distribution of dielectron masses in ultraperipheral processes
is
\begin{eqnarray}
\frac {d\sigma }{dM^2}&=&\frac {128 (Z\alpha )^4}{3M^4}
\frac {\alpha }{\sqrt {1-\frac {4m^2}{M^2}}
(1-\exp (-2\pi \alpha/\sqrt {1-\frac {4m^2}{M^2}}))}\times\nonumber \\
& &\left[(1+\frac {4m^2}{M^2}-\frac {8m^4}{M^4})
\ln \frac {1+\sqrt {1-\frac {4m^2}{M^2}}}{1-\sqrt {1-\frac {4m^2}{M^2}}}-
(1+\frac {4m^2}{M^2})\sqrt {1-\frac {4m^2}{M^2}}\right]
\ln ^3\frac {u\sqrt {s_{nn}}}{M}, \nonumber \\
\label{sM2}
\end{eqnarray}
The distribution of the relative (in $e^+e^-$ rest system) velocity $v$ is like
\begin{equation}
\frac {d\sigma }{dv^2}=\frac {16(Z\alpha )^4}{3m^2}
[(3-v^4)\ln \frac {1+v}{1-v}-2v(2-v^2)]
\frac {\alpha}{v(1-\exp (\frac {-2\pi \alpha}{v}))}
\ln^3 \frac {u\sqrt {s_{nn}(1-v^2)}}{2m}.
\label{sv2}
\end{equation}
Let us remind that the velocity $v$ is related to the velocity of the
positron $v_+$ in the electron rest system as
\begin{equation}
v^2=\frac {1-\sqrt {1-v_+^2}}{1+\sqrt {1-v_+^2}}
\label{vv}
\end{equation}
so that $v_+=2v$ at $v\rightarrow 0$ and both $v$ and $v_+$ tend to 1
in the ultrarelativistic limit. The relative velocities $v$ and $v_+$ are
the relativistic invariants represented by the Lorentz-invariant
masses $m$ and $M$.

It is shown in Fig. 1 that the cross
section of creation of unbound $e^+e^-$-pairs tends to zero at the threshold
$M=2m$ if the perturbative expression Eq. (\ref{mM}) is used. Account of the
non-perturbative SGS-factor (\ref{sgs}) in Eqs (\ref{sM2}) and (\ref{sv2})
changes the situation, drastically increasing the yield of pairs at low masses 
$M$, i.e. with small velocities $v$.

In Figs 2 and 3, we compare the yields of pairs with (curves a - (\ref{sM2}), 
(\ref{sv2})) and without (curves b - (\ref{sM})) account of the SGS-factor 
at NICA energy 11 GeV as functions of masses $M$ and velocities $v$. 

\begin{figure}
\includegraphics[width=\textwidth]{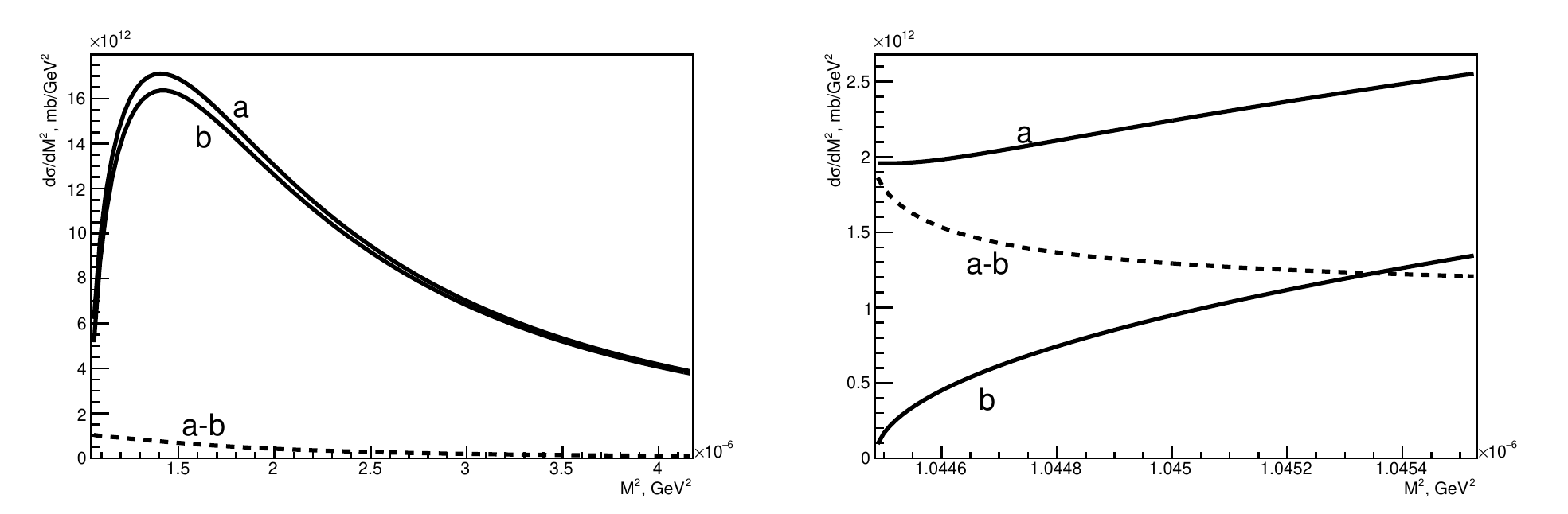}

Fig. 2. The distribution of masses of dielectrons produced in
ultraperipheral collisions at NICA energy $\sqrt {s_{nn}}$=11 GeV
with (a) and without (b) account of the SGS-factor. Their difference
(a-b) is shown by the dashed line. The region of small masses is shown
in the right-hand side at the enlarged scale. Note the factor $10^{-6}$
at the abscissa scale which reduces it to MeVs. 
\end{figure}
\begin{figure}

\centerline{\includegraphics[width=\textwidth]{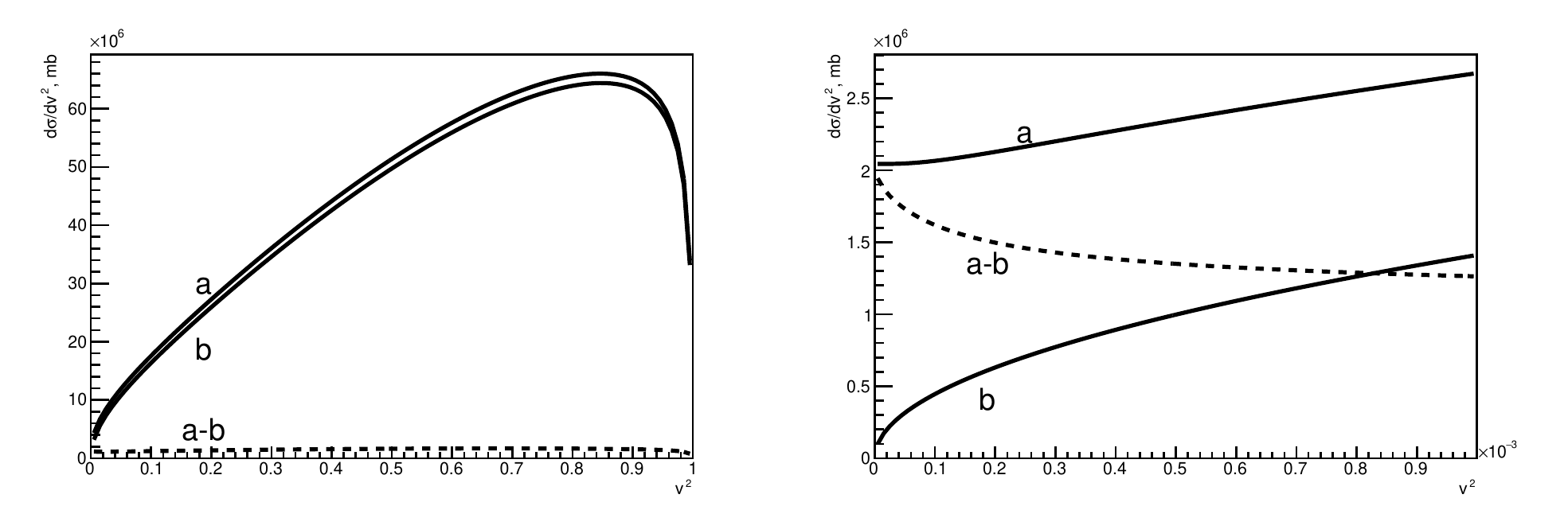}}

Fig. 3. The distribution of the relative velocities in dielectrons produced in
ultraperipheral collisions at NICA energy $\sqrt {s_{nn}}$=11 GeV
with (a) and without (b) account of the SGS-factor. Their difference
(a-b) is shown by the dashed line. The velocities for the region of small 
masses are shown in the right-hand side. Note the factor $10^{-3}$ at the 
abscissa scale.
\end{figure}

The cross sections of ultraperipheral production of unbound $e^+e^-$-pairs are
especially strongly enhanced at low masses $M$ (at low relative velocities $v$)
compared to their perturbative values (marked by b). It is clearly seen
in the righthand sides of Figs 2 and 3 which demonstrate the region
near the threshold $M=2m$. Surely, the cross section would tend to zero
at the threshold $M=2m$ due to the energy-momentum conservation laws
not fully respected by the simplified SGS-recipe. However, it must happen
in the tiny region near the threshold and can be neglected in integral
estimates.

At the same time, the
overall contribution due to the correction is not high. It amounts to
about 4.6 percents at the peak of the $M^2$ distribution and 2.5 percents
at the peak of the $v^2$ distribution. The integral contributions differ 
by 3.4 percents only.

Beside unbound pairs, parapositronia can be directly produced in two-photon 
interactions \cite{dr4}. The total cross section (\ref{e4}) is written as
\begin{equation}
\sigma_{Ps}=\frac {16Z^4\alpha ^2\Gamma }{3m^3}\ln ^3\frac {u\sqrt {s_{nn}}}{m}.
\label{e4}
\end{equation}
It is much lower than the cross section for creation of unbound pairs 
(\ref{vz}). 

The energy distribution of gamma-quanta from decays of
parapositronia produced in ultraperipheral collisions at NICA energies is
shown in Fig. 4.
\begin{figure}

\centerline{\includegraphics[width=16cm, height=14cm]{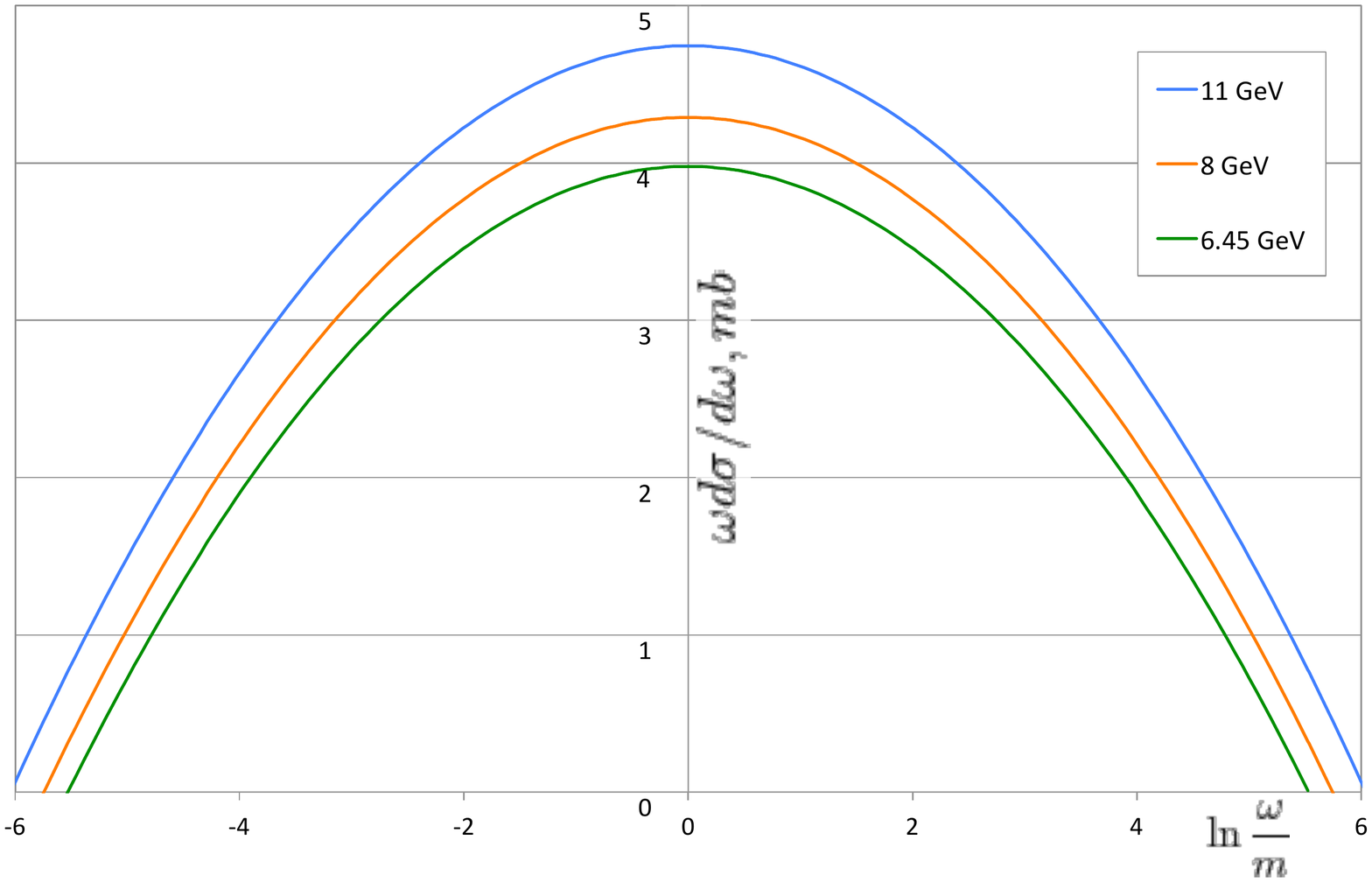}}

Fig. 4. The energy distribution of gamma-quanta from decays of
parapositronia produced in
ultraperipheral collisions at NICA energies $\sqrt {s_{nn}}$=11 GeV 
(blue, upper), 8 GeV (red, middle), 6.45 GeV (green, lower).
\end{figure}

It is obtained from Eq. (\ref{e2}) by omitting
one of the integrations there and inserting the resonance cross section
for $\sigma _{\gamma \gamma }(X)$ (see, e.g., \cite{drufn}). For Au-Au 
collisions at NICA one gets
\begin{equation}
\omega \frac {d\sigma }{d\omega }=\frac {4Z^4\alpha ^2\Gamma }{m^3}
(\ln ^2\frac {u\sqrt {s_{nn}}}{m}-\ln^2\frac {\omega }{m}), 
\label{gam}
\end{equation}
where $\Gamma \approx 5.2\cdot 10^{-15}$ GeV is the decay width of the 
parapositronium. 

This distribution (\ref{gam}) is shown in Fig. 4 as 
$\omega d\sigma /d\omega $ for three energies of NICA ranging 
from 11 GeV to 8 GeV and 6.45 GeV per nucleon. Again, the parameter $u$=0.02 is
chosen in accordance with its value obtained in Ref. \cite{vyzh}.

As expected, the photon spectra are concentrated near the electron mass
511 keV and they are rather wide. The motion of parapositronia produced in
ultraperipheral collisions at high enough energies of NICA is responsible for 
the broadened spectra in Fig. 4.

In general, the direct ultraperipheral production of parapositronia is about 
million times less effective than the creation of dielectrons as estimated 
from Eqs (\ref{vz}), (\ref{e4}). Positrons move non-relativistically relative 
to electrons in pairs with low masses. They can annihilate with high 
probability. Thus the pairs with masses near 2$m$ in Figs 2 and 3 can create
additional gamma-quanta with energies near 511 keV especially in astrophysical 
surroundings.

\section{From colliders to astrophysics}

Studies at particle accelerators demonstrate that dileptons are abundantly
produced in ultraperipheral collisions. However, the spectra of gamma-quanta 
shown in Fig. 4 are much wider than those observed from Galaxy and within
thunderstorms. It implies that kinetic energies of parapositronia are not
small in collider studies. At the same time, the small measured width of 
the galactic 511 keV line indicates that parapositronia must be almost 
at rest there. Therefore, the way to use collider data
for astrophysics and terrestrial events is not a direct one. One must point out
the reliable sources of positrons and describe the mechanism of their
cooling down to thermal energies suitable for formation of positronia at rest.
An attempt of this kind was published in Ref. \cite{dogi}. 
It was stated: "We assume the black 
hole is a source of high energy protons generated by star accretion. 
The galactic black hole could be a powerful source of relativistic protons.
Secondary positrons produced by pp collisions at energies 30 MeV are cooled
down to thermal energies by Coulomb collisions, and annihilate in the warm
neutral and ionized phases of the interstellar medium with temperatures about
several eV, because the annihilation cross-section reaches its maximum at these
temperatures.
From kinetic equations we shall show that processes of Coulomb collisions are 
effective enough to cool down these relativistic positrons and to thermalize 
them before their annihilation, which can explain the origin of the annihilation 
emission from the Galactic center." 

Here, it is demonstrated that the heavy nuclei are $Z^4$ times more effective 
in production of positrons and direct positronia than protons. Existence
of the Fe-component in cosmic rays indicates the presence of heavy nuclei
in the Galaxy. Moreover, the 
spectra of unbound pairs are much softer than those assumed in \cite{dogi}, 
especially if the Sommerfeld-Gamow-Sakharov-factor is properly accounted.
Therefore, positrons will become thermalized easier and create parapositronia
decaying to two 511 keV gamma-rays.

\section{Conclusions}

Electromagnetic fields created by fast moving charged particles are
responsible for emission of the 511 keV gamma-rays from the Galactic center 
and within the terrestrial thunderstorms. Studies at NICA collider can help
in understanding main parameters of the ultraperipheral collisions of
protons and heavy ions.

{\bf Acknowledgments}

This work was supported by the RFBR project 18-02-40131.
                      
\vspace{6pt}

The author declares no conflicts of interest.


\begin{thebibliography}{999}
\bibitem{sieg}
Siegert, T. et al., {Gamma-ray spectroscopy of Positron Annihilation in 
the Milky Way}.  
{\em AandA} {\bf 2016}, {\em 586}, A84; arXiv:1512.00325. 
\bibitem{abc}
Dwyer, J.R.; Smith, D.M.; Hazelton, B.J.; Grefenstette, B.W.; Kelley, N.A.;
Lowell, A.W.; Schaal, M.M.; Rassoul, H.K. {Positron clouds within thunderstorms}  
{\em J. Plasma Phys.} {\bf 2015}, {\em 81}, 475810405.   
\bibitem{skl}
Klein, S. {Two-photon production of dilepton pairs in peripheral heavy ion 
collisions.} 
{\em Phys. Rev. C} {\bf 2018}, {\em 97}, 054903.
\bibitem{chub1}
Chubenko, A.P.; Antonova V.P.; Kryukov S.Yu.; Piskal V.V.; Ptitsyn M.O.; 
Shepetov A.L.; Vildanova L.I.; Zybin K.P.; Gurevich A.V. {Intense 
x-ray emission bursts during thunderstorms}. {\em Phys. Lett. A} {\bf 2000}, 
{\em 275}, 90-100.
\bibitem{chub2}
Chubenko, A.P., et al. {Energy spectrum of lightning gamma emission}. 
{\em Phys. Lett. A} {\bf 2009}, {\em 373}, 2953-2958.
\bibitem{chil}
Chilingarian, A.; Mailyan B.; Vanyan L., {Recovering of the energy spectra 
of electrons and gamma rays coming from the thunderclouds}. 
{\em Atmospheric Research}, {\bf 2012}, {\em 114-115}, 1-16.
\bibitem{dr1}
Dremin, I.M., {Excess of soft dielectrons and photons.}      
{\em Universe}, {\bf 2020}, {\em 6(7)} 94; arXiv:2006.12033
\bibitem{dr2}
Dremin, I.M.; Gevorkyan, S.R.; Madigozhin, D.T., {Enhancement of low-mass dileptons 
in ultraperipheral collisions.} {\em Eur. Phys. J. C} {\bf 2021} ; arXiv:2008.13184 
\bibitem{dieh}
Diehl, R; et al, {Steady-state nucleosynthesis throughout the Galaxy.} 
arXiv:2011.06369 
\bibitem{takh}
Takhistov, V., {Positrons from Primordial Black Hole Microquasars and 
Gamma-ray Bursts.} {\em Phys. Lett. B} {\bf 2019}, {\em 789} 538.
\bibitem{isto}
Istomin, Ya.N.; Chernyshov, D.O.; Sob'yanin, D.N., {Extinct radio pulsars as 
a source of subrelativistic positrons.} 
{\em Mon. Not. R. Astron. Soc.} {\bf 2020}, {\em 498}, 2089. 
\bibitem{dogi}
Cheng, K.S.; Chernyshov, D.O.; Dogiel, V.A., {Annihilation Emission from the 
Galactic Black Hole.} {\em  Astrophys. J.} {\bf 2006}, {\em 645} 1138. 
\bibitem{cai}
Rong-Gen Cai; Yu-Chen Ding; Xing-Yu Yang; Yu-Feng Zhou,
{Constraints on a mixed model of dark matter particles and primordial 
black holes from the Galactic 511 keV line.} arXiv:2007.11804 
\bibitem{farz}
Farzan, Y.; Rajaee M., {Pico-charged particles explaining 511 keV line and 
XENON1T signal.} arXiv:2007.14421
\bibitem{lali}
Landau, L.D.; Lifshitz, E.M.
{On the production of electrons and positrons by a collision of two particles}.
{\em Physikalische Zeitschrift der Sowjetunion} {\bf 1934},~{\em 6},~244.
\bibitem{froi}
Froissart, M., {Asymptotic Behavior and Subtractions in the Mandelstam 
Representation.} 
{\em Phys. Rev.} {\bf 1961}, {\em 123} 1053.
\bibitem{dr3}
Dremin, I.M., {Ultraperipheral vs ordinary nuclear interactions.}      
{\em Universe}, {\bf 2020}, {\em 6(1)} 4; arXiv:1910.09838
\bibitem{ijmp}
Dremin, I.M., {Thresholds of ultraperipheral processes.} 
{\em Int. J. Mod. Phys. A} {\bf 2020}, {\em 35}, 2050087.
\bibitem{wei}
Weizs\"{a}cker, C.F.V. 
{Radiation emitted in collisions of very fast electrons}.
{\em Zeit. Phys.} {\bf 1934}, ~{\em 88},~612--625.
\bibitem{wil}
Williams, E.J. {Nature of the high energy particles of penetrating radiation 
and status of ionization and radiation formulae.
\emph{Phys. Rev.} \textbf{1934},~\emph{45},~729--730.}
\bibitem{rac}
Racah, G., {Sulla Nascita di Coppie per Urti di Particelle Elettrizzate.}
{\em Nuovo Cim.}, {\bf 1937}, {\em 14}, 93.
\bibitem{bgms}
Budnev, V.M.; Ginzburg, I.F.; Meledin, G.V.; Serbo, V.G., {The two-photon particle 
production mechanism. Physical problems. Applications. Equivalent photon 
approximation.} 
{\em  {Phys. Rep. C}} {\bf 1975},~{\em 15},~181--282. 
\bibitem{drufn}
Dremin, I,M., {Ultraperipheral nuclear interactions.}
{\em Phys. Usp.}, {\bf 2020}, {\em 190}, 811. 
\bibitem{blp}
Berestetsky, V.B.; Lifshitz, E.M.; Pitaevsky, L.P.,
\emph{Kvantovaya Electrodinamika}
(Fizmatlit: Moscow, Russia, 2001)
\bibitem{vyzh}
Vysotsky, M.I.; Zhemchugov, E.V., {Equivalent photons in proton-proton and 
ion-ion collisions at the Large Hadron Collider.}
\emph{Phys. Usp.} {\bf 2019},~{\em 189},~975--984.
\bibitem{brwh}
Breit, G.; Wheeler, J.A., {Collision of Two Light Quanta}
{\em Phys. Rev.} {\bf 1934}, {\em 46}, 1087.
\bibitem{llqm}
Landau, L.D.; Lifshitz, E.M., {\em Kvantovaya Mechanika,
 Nerelyativistskaya Teoriya}, 2 izdanie, (134.11) (Fizmatlit, Moscow ,1963)
 (translated in: {\em Quantum Mechanics} (Pergamon Press, Oxford, 1977))
\bibitem{som}
Sommerfeld, A., {\em Atombau und Spectrallinien}, (F. Vieweg und Sohn,  
Brunswick, Deutschland, 1921)
\bibitem{gam}
Gamow, G., {Zur Quantentheorie des Atomkernes.} 
{\em Zeit. Phys.}, {\bf 1928}, {\em 51}, 204.
\bibitem{somm}
Sommerfeld, A., {\em Ann. Phys. (Leipz.)}, {\bf 1931}, {\em 403}, 257.
\bibitem{sakh}
Sakharov, A.D., {Interaction of the electron and the positron in pair 
production.}
{\em Zh. Eksp. Teor. Fiz.}, {\bf 1948}, {\em 18}, 631
(Reprinted in {\em Sov. Phys. Usp.}, {\bf 1991}, {\em 34}, 375).
\bibitem{baier}
Baier, V.N.; Fadin, V.S., {Coulomb interaction in the final state.}
{\em Sov.Phys. JETP}, {\bf 1970}, {\em 30}, 127.
\bibitem{ieng}
Iengo,R., {Sommerfeld enhancement: general results from field theory diagrams.}
{\em JHEP}, {\bf 2009}, {\em 05}, 024; arXiv:0902.0688.                         
\bibitem{cass}
Cassel, S., {Sommerfeld factor for arbitrary partial wave processes.}
{\em J. Phys. G}, {\bf 2010}, {\em 37}, 105009; arXiv:0903.5307.
\bibitem{arko}
Arbuzov, A.B.; Kopylova, T.V., {On relativization of the Sommerfeld-Gamow-Sakharov
factor.} {\em JHEP}, {\bf 2012}, {\em 04}, 009; arXiv:1111.4308.
\bibitem{dr4}
Dremin, I.M., {Geometry of ultraperipheral nuclear collisions.}
{\em Int. J. Mod. Phys. A}, {\bf 2019}, {\em 34}, 1950068; arXiv:1903.12377 

\end{thebibliography}
\end{document}